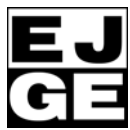

# Institutional quality and the wealth of autocrats

**Christopher J. Boudreaux**[a,][*], **Randall Holcombe**[b]

[a] Texas A&M International University, U.S.A.

[b] Department of Economics, Florida State University, U.S.A.

**\* Corresponding author at**: Texas A&M International University, Laredo, TX 78041. chrisboudreaux1@gmail.com

**Abstract.** One frequently given explanation for why autocrats maintain corrupt and inefficient institutions is that the autocrats benefit personally even though the citizens of their countries are worse off. The empirical evidence does not support this hypothesis. Autocrats in countries with low-quality institutions do tend to be wealthy, but typically, they were wealthy before they assumed power. A plausible explanation, consistent with the data, is that wealthy individuals in countries with inefficient and corrupt institutions face the threat of having their wealth appropriated by government, so have the incentive to use some of their wealth to seek political power to protect the rest of their wealth from confiscation. While autocrats may use government institutions to increase their wealth, autocrats in countries with low-quality institutions tend to be wealthy when they assume power, because wealthy individuals have the incentive to use their wealth to acquire political power to protect themselves from a potentially predatory government.

## 1. Introduction

An important question in economic development is why autocrats in poor countries maintain low-quality economic institutions when a substantial body of literature concludes that specific institutional improvements can produce higher economic growth rates and higher per capita incomes. One common answer is that while the general population may be better off with higher-quality economic institutions, the political elite is better off maintaining more oppressive economic institutions that benefits it, even though those institutions make their countries poorer. Referring to the poor states of Africa, Ayittey (2008: 168) says, "African despots are loath to relinquish control or power. They would rather destroy their economies and countries than give up economic and political power. This power allows them to allocate or extract resources to build personal fortunes and to dispense patronage to buy political support." Acemoglu and Robinson (2006: 115) say it is "…because superior institutions and technologies may reduce their political power and make it more likely that they will be replaced." Acemoglu and Robinson (2013) emphasize the importance of taking into account both economic and political factors when looking at the potential for economic reform, and offer examples that illustrate the peril of not doing so.

If autocrats benefit from maintaining low-quality economic institutions, an important question is what benefit autocrats actually receive. The few systematic empirical studies that have been done have not found evidence that the political elite receive any benefit from low-quality economic institutions. This paper uses a small dataset that looks directly at autocrat wealth to see if it is positively impacted, and finds no evidence that it is. The evidence indicates that if anything, autocrats benefit from overseeing institutional improvements in their countries, and one would conjecture that if this research is correct, autocrats themselves would be in a good position to recognize the benefits they could reap from improving institutions. A plausible explanation for why they do not improve institutions in their countries is that institutional change is difficult to implement and cannot be accomplished by a single individual, even if that individual wants to improve institutional quality and has substantial political power.

The casual observation that poor countries tend to be ruled by wealthy autocrats may have led researchers to consider the wrong question. Wealthy autocrats tend to be wealthy when they assume power, so a more appropriate question is why wealthy individuals in countries with inefficient and corrupt institutions seek political power when the wealthy do not typically do so in countries with higher-quality institutions. The wealthy in the United States, Western Europe, and Japan, for example, tend to focus on their economic activities rather than seek political office.[1] The observation of wealthy autocrats in poor countries is consistent with the hypothesis that when institutional quality is poor, the wealthy use some of their wealth to seek political power to protect the rest of their wealth from confiscation by the political elite. Wealthy autocrats hold political power in poor counties not because those institutions enable them to enhance their wealth, but rather because those governments tend to be more predatory, so the wealthy seek political power to protect their wealth from government predation.

## 2. Evidence on the Benefits to Autocrats

Despite the conventional wisdom that autocrats receive personal benefits from maintaining low-quality economic institutions at the expense of the well-being of their citizens, there is relatively little evidence to support that conclusion. As noted in the introduction, Ayittey (2008) offers anecdotal evidence regarding African dictatorships. Acemoglu and Robinson (2006) develop a theoretical framework in which policies that can produce economic development for the general population can also impose costs on political elites, causing them to choose to inhibit economic development for their own benefit. They offer historical examples of Czarist Russia and the nineteenth century Austro-Hungarian Empire to support their case, but also show, using the United States, Germany, and Britain as examples, that political elites sometimes implement development-enhancing institutions. Acemoglu and Robinson (2013) look at destabilizing economic reforms in Ghana and Sierra Leone to show how improving economic institutions can work to the detriment of political elites. This evidence is anecdotal: a theoretical framework with case studies that provide examples consistent with the model.

---

[1] Donald Trump is a notable exception here, but Trump is different from most other wealthy Americans in that he was a television personality prior to running for office, a profession that rewards a high-profile public image. Unlike President Trump, most wealthy individuals try to keep a low profile and remain out of the public eye rather than seeking personal recognition, as Trump did well before he ran for public office.

Anecdotes can also be found that lean the other way. Both Augusto Pinochet in Chile and Park Chung-hee in South Korea were autocrats who held power for 16 years in their countries and implemented substantial economic reform, which provided benefits to both the autocrats and their citizens.[2] Examples can be cited both for countries in which autocrats ruled countries that remained poor and with low-quality institutions and cases in which autocrats improved institutional quality and oversaw institutional improvement and increases in income. If better institutions create wealth, there should be a way to distribute that wealth so that everyone benefits, although Rodrik (2014) notes that it can be difficult to design institutional reforms that guarantee that everyone will share in the gains. Anecdotes can be chosen to support various conclusions.

Evidence on the issue is mostly anecdotal and we are aware of only two studies that look at a cross-section of countries for evidence that political elites benefit personally from maintaining low-quality economic institutions that reduce the standard of living for their citizens. Holcombe and Rodet (2012) use government spending as a proxy for benefits to political elites under the assumption that while those elites cannot appropriate all that government spends, they control who gets the benefits of that spending. They look at a set of more than 100 countries with all types of governments – not just autocracies – to see if lower-quality economic institutions increase the size of government, both in real per capita terms and as a share of total income. They find that better economic institutions are associated with a larger government share, which suggests that political elites could benefit from improved economic institutions. Political elites have control over more resources when economic institutions are improved, calling into question the conventional wisdom that political elites benefit from maintaining low-quality economic institutions.

Looking at the claim of Acemoglu and Johnson (2006) that institutional reforms may destabilize the incumbent regime, Holcombe and Boudreaux (2013) examine a dataset of 80 autocrats and find that the greater the increase in the quality of economic institutions under their tenure, the longer their tenure, indicating that autocrats can increase their hold on power by improving institutional quality. The few studies that analyze a dataset of multiple countries rather than anecdotal evidence and case studies conclude that autocrats can improve their control over resources and their hold on power by implementing higher-quality economic institutions, calling into question the hypothesis that autocrats maintain poor institutions because they benefit at the expense of their citizens.

---

[2] Another possibly relevant example is Mikheil Saakashvili's presidency of the Republic of Georgia from 2004-2013. Although Saakashvili was elected, Georgia was poor and its government was corrupt and inefficient when he came to power. Substantial economic reforms led Georgia to prosper relative to its economic status prior to his presidency and when compared with other former Soviet republics in the Caucuses. As president, Saakashvili was also accused of appropriating property without compensation and cronyism during his term in office. This is an example of a political leader who improved economic institutions both for the benefit of his citizens and himself and his cronies as an elected leader.

This study examines additional evidence on the relationship between autocrat wealth and the quality of economic institutions. The evidence here is more direct than in previous studies, although it is from a limited dataset on the wealth of 35 autocrats. There is no comprehensive data available on the wealth of autocrats, so this study relies on data collected from various sources on that wealth. Recognizing the limitations of the data, the findings here are conclusive. Rather, they offer some additional evidence that calls into question the hypothesis that autocrats benefit personally from maintaining low-quality economic institutions.

The results that follow are consistent with the hypothesis that in countries with low-quality institutions, wealthy individuals are attracted to political power, perhaps to protect their wealth from predation that can occur in countries that have a poor record of protecting property rights and establishing rule of law.

## 3. Data on the Wealth of Autocrats and Institutional Quality

Table 1 lists 35 autocrats along with their countries, the time periods in which they were in power, and their wealth. Data sources are listed in Appendix 1. The dataset is limited because these were the only autocrats for which we were able to find estimates of their wealth. Wealth information was found for leaders of countries who ruled in 2010. In a few cases, leaders from previous years were also included in the analysis. Khaleda Zia of Bangladesh exited office in 2006, Mobutu Sese Seko of the Democratic Republic of Congo exited in 1997, and Mohamed Suharto of Indonesia exited in 1998. For these leaders, wealth estimates are adjusted for inflation and stated in 2010 U.S. dollars.

Included in the list are several monarchs who have limited power over their governments but who have retained power that is largely symbolic. While they would not ordinarily be classified as autocrats, they make a good comparison group as wealthy heads of state in countries that have relatively high-quality economic and political institutions. This allows a comparison of the effects of institutional quality that includes a wider range of institutions, rather than just low-quality institutions.

The quality of economic institutions was measured using the Fraser Institute's Economic Freedom of the World (EFW) Index, updated annually by Gwartney, Lawson, and Hall (2012). The EFW index is frequently used as a measure of the quality of economic institutions. It is designed to quantify the level of economic freedom that exists in a country, and is deliberately designed to quantify the quality of economic institutions but to leave out measures of political freedom, such as civil liberties or democratic government. Thus, it is a good measure of institutional quality to use for the examination of the hypothesis that autocrats can increase their wealth by imposing economically unproductive institutions.

**Table 1.** List of leaders.

| Country | Leader | years | $ Wealth † |
|---|---|---|---|
| Angola | Jose Eduardo Dos Santos | 1979-Present | 31 billion |
| Azerbaijan | Ilham Aliyev | 2003-Present | 500 million |
| Bahrain | Hamad ibn isa al Khalifa | 2002-Present | 5.44 billion |
| Bangladesh | Khaleda Zia | 1991-1996; 2001-2006 | 220 million |
| Cameroon | Paul Biya | 1982-Present | 200 million |
| China | Wen Jiabao | 2003-Present | 2.7 billion |
| Colombia | Juan Manuel Santos | 2010-Present | 215 million |
| Dem. Rep. of Congo | Mobutu Sese Seko | 1970-1997 | 6.9 billion |
| Egypt | Hosni Mubarek | 1981-2011 | 40 billion |
| Ethiopia | Girma Wolde-Giorgis | 2001-Present | 1.5 million |
| Gabon | Ali Bongo Ondimba | 2009-Present | 1 billion |
| Indonesia | mohamed suharto | 1967-1998 | 21.8 billion |
| Iran | Ali Khamenei | 1981-Present | 30 billion |
| Kazakhstan | Nursultan Nazarbayev | 1991-Present | 1.11 billion |
| Kuwait | H.H Sheikh Sabah Al-Ahmad Al-Jaber Al-Sabah | 2006-Present | 14.23 billion |
| Montenegro | Milo Đukanović | 2003-2006; 2008-2010 | 15 million |
| Morocco | Mohammad IV | 1999-Present | 2.5 billion |
| Myanmar | Than Shwe | 1992-2011 | 4 billion |
| Netherlands | Beatrix | 1980-2013 | 200 million |
| Nigeria | Goodluck Jonathan | 2010-Present | 2.8 million |
| Norway | Harald V | 1991-Present | 17.3 million |
| Oman | Qaboos Bin Said Al Said | 1970-Present | 9.33 billion |
| Pakistan | Asif Ali Zardari | 2008-Present | 1.8 billion |
| Philippines | Benigno Noynoy Aquino III | 2010-Present | 1.3 million |
| Qatar | Hamad bin Khalifa | 1995-Present | 10.88 billion |
| Saudi Arabia | Abdullah Bin Abdulaziz | 2005-Present | 21 billion |
| Spain | Juan Carlos I | 1975-2014 | 5 billion |
| Sweden | Carl XVI Gustaf | 1973-Present | 42 million |
| Syria | Bashar Al-Assad | 2000-Present | 1 billion |
| Thailand | Bhumibol Adulyadej | 1946-Present | 30 billion |
| Tunisia | Zine El Abidine Ben Ali | 1987-2011 | 7.8 billion |
| U.A.E. | Sheikh Khalifa bin Zayed Al Nahyan | 2004-Present | 21 billion |
| Uganda | Yoweri Musevini | 1986-Present | 1.7 billion |
| United Kingdom | Elizabeth II | 1952-Present | 450 million |
| Zimbabwe | Robert Mugabe | 1987-2017 | 10 billion |

**Note** - † wealth amounts ending before 2010 are converted into 2010 dollars.

Literature reviews by Berggren (2003) and De Haan, Lundstrom, and Sturn (2006) show that countries with higher-quality economic institutions as measured by the EFW index, have higher per capita incomes, and countries that improve their economic freedom as measured by the index have higher rates of economic growth. Subsequent studies, such as Faria and Montesinos (2009), have reaffirmed the positive impact of market institutions. Countries with higher quality institutions, as measured by the EFW index, are more prosperous. This makes the EFW index the ideal measure of the quality of economic institutions for present purposes, because of the many studies that have shown it is positively correlated with per capita income and income growth. While there is little doubt that higher-quality economic institutions improve the economic well-being of a country, the question examined here is whether autocrats who maintain low-quality institutions to the detriment of their citizens receive personal benefits in the form of higher personal wealth. It is a more direct measure of personal benefit than was used in any of the previous studies.

## 4. Institutional Quality and the Wealth of Autocrats

The hypothesis that autocrats maintain low-quality institutions to increase their wealth is tested directly in an OLS model by looking at the correlation between autocrat wealth and institutional quality. The dependent variable is the log of the wealth of each political leader, W, and the major independent variables are measures of the quality of economic and political institutions at the beginning of the leaders' tenure and the average annual change in institutional quality during their tenures. Economic institutions are measured by the EFW score at the beginning of the leaders' tenures, InitialEFW, and the average annual change in the EFW measure, EFWRate, over the leaders' tenures. EFWRate is calculated by subtracting InitialEFW from the final EFW value and dividing by the number of years the leader held power. Four different measures are used for the quality of political institutions, which are measured the same way. Initial political rights are represented as InitialPL, and the change over the autocrats' tenures is PLRate. The measures used are the Polity 2 index, the Freedom House index of political freedom, and the two individual components of the Freedom House index: Political Rights, and Civil Rights.

InitialEFW and InitialPR are variables outside the control of the autocrat, but they could influence the autocrat's wealth. EFWRate and PLRate are variables that, according to the hypothesis being examined, are under the autocrat's control. The hypothesis predicts that autocrats choose poor institutions for their own benefit, but they could not choose those institutions if institutional quality is outside their control. The hypothesis being examined assumes a direction of causation that runs from EFWRate and PLRate to autocrat wealth.

Other factors could also affect autocrat wealth. Most obviously, people in wealthier countries tend to be wealthier, so per capita income when the autocrat comes to power, InitialPCI, and the change in per capita income over the autocrat's tenure, PCIRate, are included. Some scholars, such as Diamond (1997) and Sachs (2001), have suggested that natural resource endowments can affect incomes and institutional quality, so a measure of resource rents, R, is included in the empirical model. Easterly (2006) and Coyne (2013) have suggested foreign aid is often counterproductive to the countries that receive it, although it might benefit a country's political elite, prompting the inclusion of foreign aid, F, in the empirical model. Another geographic factor often associated with a state's economic well-being is distance from the equator, D. Following Holcombe and Boudreaux (2013), the length of tenure, T, of an autocrat might be related to the autocrat's wealth, and a binary variable, M, is included which is 1 for autocrats who are monarchs and 0 otherwise.[3] Appendix 2 shows variable definitions and sources. The empirical specification is

$$W = \alpha + \beta_1 \text{InitialEFW} + \beta_2 \text{EFWRate} + \beta_3 \text{InitialPCI} + \beta_4 \text{PCIRate} + \beta_5 R + \beta_6 F + \beta_7 D + \beta_8 T + \beta_9 M + \beta_{10} \text{InitialPI} + \beta_{11} \text{PIRate} + \varepsilon.$$

The empirical results for eight different empirical specifications are reported in Table 2.

The specifications differ only in that four of them include only EFWRate, not, InitialEFW, and different measures are used to measure the differences in political institutions. In the eight different specifications, InitialEFW is never statistically significant at the 10 percent level or better, while EFWRate is positive and statistically significant at the 10 percent level in three of the four specifications which omit InitialEFW. The results in Table 2 indicate that the quality of economic institutions at the time an autocrat takes office are unrelated to autocrat wealth, but there is weak evidence that wealth is higher when the quality of economic institutions improves over the autocrat's tenure. Consistent with previous studies, those results show that if anything, autocrats benefit from improvements in the quality of economic institutions during their tenures.

Initial per capita income is statistically significant at the 10 percent level in half of the regressions, and the autocrat's length of tenure is statistically significant in all of them. Autocrats who rule over higher-income countries tend to have higher levels of wealth, and autocrats who retain power longer tend to have higher wealth. The direction of causation of the tenure variable is uncertain. Wealthier autocrats might have an advantage in maintaining power, but it also may be that longer tenures allow more time to accumulate wealth. The wealth variable measures wealth at the end of the autocrat's tenure, and data limitations do not allow calculating the change in the autocrat's wealth over the autocrat's tenure. Perhaps the Tenure variable is significant because autocrats with longer tenures tend to come to power at a younger age. To check for this, the age that autocrats assumed power was substituted for Tenure, and the age variable was significant. However, when both age and Tenure were included in the same regressions, Tenure remained significant but age did not. Thus, we report the specifications than include only the Tenure variable.

---

[3] All specifications were also run without the monarchy variable and the institutional variables were qualitatively the same, so for space considerations the paper does not report the results without the monarchy variable.

**Table 2.** Effect of Economic and Political Institutions on the Wealth of Leaders.

|  | Log wealth | | | | | | | |
|---|---|---|---|---|---|---|---|---|
|  | (1) | (2) | (3) | (4) | (5) | (6) | (7) | (8) |
| **Initial EFW** |  | -0.357 |  | -0.275 |  | -0.510 |  | -0.536 |
|  |  | (-1.08) |  | (-0.69) |  | (-1.33) |  | (-1.27) |
| **EFW rate** | 4.866 | 1.753 | 8.139+ | 5.578 | 9.924+ | 5.138 | 9.713+ | 4.501 |
|  | (1.19) | (0.42) | (1.74) | (0.98) | (1.80) | (0.88) | (1.86) | (0.74) |
| **Initial PCI†** | 0.0220 | 0.0410 | 0.0688* | 0.0817* | 0.0668 | 0.0912* | 0.0548 | 0.0813+ |
|  | (0.86) | (1.32) | (2.17) | (2.13) | (1.70) | (2.09) | (1.38) | (1.79) |
| **PCI rate†** | -0.0864 | -0.0647 | -0.128 | -0.0992 | -0.206 | -0.168 | 0.0907 | 0.0834 |
|  | (-0.27) | (-0.19) | (-0.51) | (-0.36) | (-0.60) | (-0.46) | (0.22) | (0.19) |
| **Resource rents** | 0.0317 | 0.0314 | 0.00305 | 0.00442 | 0.0104 | 0.0120 | 0.0184 | 0.0215 |
|  | (1.30) | (1.31) | (0.15) | (0.20) | (0.41) | (0.46) | (0.64) | (0.72) |
| **Foreign aid** | -8.823 | -11.30 | -9.652 | -11.33 | -8.410 | -12.12 | -9.650 | -13.18 |
|  | (-1.23) | (-1.47) | (-0.81) | (-0.88) | (-0.77) | (-1.02) | (-0.77) | (-0.98) |
| **Distance†** | 0.101 | 0.111 | 0.0573 | 0.0722 | 0.0280 | 0.0496 | -0.0300 | -0.0121 |
|  | (0.47) | (0.50) | (0.29) | (0.35) | (0.13) | (0.22) | (-0.14) | (-0.05) |
| **Tenure** | 0.0433+ | 0.0425+ | 0.0685** | 0.0670** | 0.0591* | 0.0565* | 0.0941** | 0.0879** |
|  | (2.00) | (1.98) | (2.89) | (2.99) | (2.39) | (2.49) | (3.10) | (3.18) |
| **Initial Polity2** | -0.181* | -0.177* |  |  |  |  |  |  |
|  | (-2.49) | (-2.37) |  |  |  |  |  |  |
| **Polity2 rate** | 7.728** | 7.206* |  |  |  |  |  |  |
|  | (2.80) | (2.57) |  |  |  |  |  |  |
| **Monarchy** | 1.064 | 1.026 | 0.521 | 0.545 | 0.650 | 0.655 | 0.702 | 0.754 |
|  | (1.38) | (1.33) | (0.50) | (0.52) | (0.68) | (0.69) | (0.59) | (0.66) |
| **Initial Freedom house** |  |  | -1.140*** | -1.109*** |  |  |  |  |
|  |  |  | (-5.62) | (5.01) |  |  |  |  |
| **Freedom rate** |  |  | -11.90+ | -10.49 |  |  |  |  |
|  |  |  | (2.03) | (1.57) |  |  |  |  |
| **Initial Political rights** |  |  |  |  | -0.926*** | -0.879** |  |  |
|  |  |  |  |  | (3.78) | (3.38) |  |  |
| **Political rights rate** |  |  |  |  | 1.906 | 1.511 |  |  |
|  |  |  |  |  | (-0.19) | (-0.15) |  |  |
| **Initial civil rights** |  |  |  |  |  |  | -1.024** | -0.942* |
|  |  |  |  |  |  |  | (3.07) | (2.61) |
| **Civil rights rate** |  |  |  |  |  |  | -7.338 | -7.686 |
|  |  |  |  |  |  |  | (-1.11) | (-1.24) |
| **Constant** | -2.836* | -0.691 | -7.579*** | -5.864* | -6.809** | -3.572 | -7.552** | -3.972 |
|  | (-2.38) | (-0.35) | (-5.02) | (-2.26) | (-3.26) | (-1.25) | (-3.47) | (-1.16) |
| *Adj. $R^2$* | 0.56 | 0.56 | 0.55 | 0.54 | 0.46 | 0.47 | 0.37 | 0.37 |

**Note**. N=35 observations. t statistics in parentheses. + $p<0.10$ * $p<0.05$ ** $p<0.01$ *** $p<0.001$

One possible area of concern in the list of autocrats used here is that they represent governments with vastly different government structures. The list includes the Queen Elizabeth II of Great Britain and Northern Ireland and former President Robert Mugabe of Zimbabwe, for example. The inclusion of these measures of political institutions should mitigate those differences by taking into account the political structure. As a robustness check, the regressions were run without Elizabeth II, and without Assad of Syria, and the results are essentially unaffected.[4] Monarchs were also separated out with a binary variable, and Table 2 shows that the monarchy variable was never statistically significant.

The quality of political institutions is controlled for in the same way as economic institutions. The first two regressions use the Polity2 index to control for political institutions. The index rates the degree of democratic control of government, with values ranging from -10 for the least democratic institutions to 10 for the most democratic. In the first two regressions, the negative sign on Initial Polity2 shows that the less democratic are the initial political institutions, the higher is the wealth of the autocrat. The positive sign on Polity2 Rate shows that movements toward more democratic institutions during the autocrat's tenure are associated with higher autocrat wealth. While the quality of economic institutions appears to have weak effects on autocrat wealth, as political institutions do. The less democratic a state's institutions when the autocrat assumes power, the more wealthy the autocrat, and the more the country moves toward democratic institutions during the autocrat's tenure, the higher the wealth of the autocrat. This suggests a benefit to the autocrat from moving toward more democratic institutions.

The remaining six regressions use the Freedom House index of political freedom as a robustness check on this result. The Freedom House index measures both civil rights and political rights—the degree to which citizens have control over their governments—as components of political freedom. Regressions 3 and 4 replace the Polity2 index with the Freedom House index of political freedom, and while the initial level of political freedom is negative and significant, consistent with the Polity2 results, Freedom Rate, which is the average annual change in political freedom over the autocrat's tenure is not statistically significant. The Freedom House index looks at political institutions more broadly than Polity2, and the political rights component is more consistent with the democratic control measured by Polity2 than is the entire index. Regressions 5 and 6 include only the political rights component of the Freedom House index. The results are similar to those using the entire index. Initial political rights show up negative and significant, as with the entire index, the Political Rights rate is not statistically significant. Regressions 7 and 8 include only the civil rights component of the Freedom House index, and yield the same qualitative result.

---

[4] Assad was deleted as a robustness check because there is some question about the actual amount of his wealth.

**Table 3.** Effect of Economic and Political Institutions on the Wealth of Leaders (Ordered Probit model).

| | Wealth | | | | | | | |
|---|---|---|---|---|---|---|---|---|
| | (1) | (2) | (3) | (4) | (5) | (6) | (7) | (8) |
| **Initial EFW** | -0.0967 | 0.404 | -0.217 | 0.319 | -0.203 | 0.390 | -0.227 | 0.277 |
| | (-0.54) | (1.56) | (-1.14) | (1.21) | (-1.14) | (1.29) | (-1.21) | (1.03) |
| **EFW rate** | | 20.64** | | 22.08*** | | 26.95*** | | 22.08*** |
| | | (2.93) | | (3.68) | | (3.32) | | (3.58) |
| **Initial PCI†** | 0.0380+ | 0.0135 | 0.0870** | 0.0681* | 0.0716** | 0.0712* | 0.0774** | 0.0543* |
| | (1.95) | (0.50) | (2.95) | (2.33) | (2.60) | (2.48) | (2.92) | (1.97) |
| **PCI rate†** | -0.302 | -0.125 | -0.313 | -0.174 | -0.449 | -0.061 | -0.431 | -0.164 |
| | (-0.91) | (-0.38) | (-0.93) | (-0.54) | (-1.12) | (-1.02) | (-1.14) | (-0.48) |
| **Resource rents** | 0.0102 | 0.0291 | 0.00521 | 0.0208 | 0.00248 | 0.0202 | 0.0112 | 0.0252 |
| | (0.62) | (1.57) | (0.34) | (1.28) | (0.16) | (1.25) | (0.67) | (1.53) |
| **Foreign aid** | -8.657+ | -7.829 | -8.866+ | -8.276 | -9.496+ | -9.218 | -9.047+ | -9.018 |
| | (-1.78) | (-1.26) | (-1.66) | (-1.19) | (-1.73) | (-1.23) | (-1.73) | (-1.25) |
| **Distance†** | -0.00798 | -0.149 | -0.0330 | -0.150 | -0.0255 | -0.185 | -0.0744 | -0.180 |
| | (-0.07) | (-1.07) | (-0.25) | (-1.08) | (-0.21) | (-1.28) | (-0.58) | (-1.45) |
| **tenure** | 0.0150 | 0.0221 | 0.0380** | 0.0458** | 0.0212+ | 0.0313* | 0.0309* | 0.0452** |
| | (1.19) | (1.60) | (2.66) | (2.63) | (1.65) | (2.00) | (2.16) | (2.95) |
| **Initial polity2** | -0.0900+ | -0.0982* | | | | | | |
| | (-1.94) | (-2.12) | | | | | | |
| **Polity2 rate** | 2.528 | 3.189* | | | | | | |
| | (1.58) | (2.03) | | | | | | |
| **Monarchy** | 1.006* | 1.482* | 1.164** | 1.410** | 0.895* | 1.390** | 1.156** | 1.323** |
| | (2.26) | (2.51) | (2.61) | (2.92) | (2.21) | (3.05) | (2.71) | (2.83) |
| **Initial Freedom house** | | | -0.567*** | -0.672*** | | | | |
| | | | (-3.45) | (4.30) | | | | |
| **Freedom rate** | | | 5.073 | 2.386 | | | | |
| | | | (-1.47) | (-0.71) | | | | |
| **Initial Political rights** | | | | | -0.392** | -0.572*** | | |
| | | | | | (2.81) | (4.06) | | |
| **Political rights rate** | | | | | 1.323 | 3.715 | | |
| | | | | | (-0.32) | (-0.70) | | |
| **Initial Civil rights** | | | | | | | -0.409* | -0.547** |
| | | | | | | | (2.32) | (3.06) |
| **Civil rights rate** | | | | | | | -8.781* | -6.811 |
| | | | | | | | (-2.40) | (-1.24) |
| **LR X²** | 27.94 | 40.18 | 32.10 | 44.36 | 25.92 | 43.57 | 25.22 | 34.25 |

**Note**. N=35 observations. t statistics in parentheses. + p<0.10 * p<0.05 ** p<0.01 *** p<0.001. † in thousands. Dependent variable is ordered in four categories.

In all regressions, whether the Polity2 index or variants of the Freedom House index are used to quantify the quality of political institutions, lower-quality political institutions are associated with higher levels of autocrat wealth. Autocrat wealth tends to be higher when autocrats come to power in countries that are less democratic, where citizens have fewer political rights and lower levels of civil rights. Assuming that political freedom and more democratic control of government are good things, the lower the quality of political institutions when an autocrat assumes power, the higher will be the autocrat's wealth. Regardless of how the quality of those political institutions is measured, the coefficients are always significant at the 1 percent level.

As a robustness check on these results, Autocrat Wealth was respecified into four categories: 1: those with wealth under $100 million; 2: wealth between $100 million and $1 billion; 3: wealth between $1 billion and $10 billion; and 4: wealth greater than $10 billion. Ordered probits were run using the same eight specifications as in Table 2, and those results appear in Table 3. The results are mostly unchanged, but with two differences worth noting. The Polity2 Rate is not significant in the ordered probit, but more noteworthy, EFWRate is strongly significant and positive in all four specifications in which it is included.

Autocrats are wealthier in countries that have improving economic institutions under their rule. Consistent with Holcombe and Rodet (2012) and Holcombe and Boudreaux (2013), this suggests that an increase in the quality of economic institutions makes autocrats better off, which calls into question the conventional wisdom that autocrats benefit from maintaining low-quality institutions. Consistent with the results in Table 2, the measures of initial political institutions all remain negative and highly statistically significant. Also note that, in contrast to Table 2, the monarchy variable is positive and statistically significant in all of the regressions in Table 3. While this indicates that monarchs tend to be wealthier than other autocrats, the institutional variables are qualitatively the same in both tables.

Table 4 shows a correlation matrix with all of the economic and political institutions variables. The Table shows a positive, although weak, correlation between the Initial EFW and all of the initial political variables.[5] Because the measured quality of economic institutions and the measured quality of political institutions are positively correlated, what appears to be the effect of economic institutions may actually be the result of political institutions. People have observed wealthy autocrats overseeing countries with low-quality economic institutions and inferred that those autocrats benefit from maintaining low-quality institutions, but this evidence points to a different conclusion: countries with lower-quality political institutions tend to attract wealthier autocrats. The positive correlation between the quality of political and economic institutions means that those wealthy autocrats will also be ruling countries with low-quality economic institutions. Autocrat wealth is negatively correlated with the quality of political institutions when they assume power.

---

[5] Also note that there tends to be a negative correlation between the initial variables and the rate variables. This may indicate convergence in institutional quality, although the correlation coefficients are small enough in magnitude that one would hesitate to draw firm conclusions from them.

**Table 4**. Correlation matrix.

|  | Initial efw | Efw rate | Initial Polity 2 | Polity2 rate | Initial freedom house | Freedom rate | Initial civil rights | Civil rights rate | Initial political rights | Political rights rate |
|---|---|---|---|---|---|---|---|---|---|---|
| **Initial efw** | 1 | | | | | | | | | |
| **Efw rate** | -0.530 | 1 | | | | | | | | |
| **Initial Polity 2** | 0.179 | -0.140 | 1 | | | | | | | |
| **Polity2 rate** | -0.435 | 0.257 | -0.277 | 1 | | | | | | |
| **Initial freedom house** | 0.294 | -0.154 | 0.784 | -0.216 | 1 | | | | | |
| **Freedom rate** | 0.125 | 0.156 | 0.246 | 0.021 | -0.072 | 1 | | | | |
| **Initial civil rights** | 0.314 | -0.167 | 0.752 | -0.165 | 0.932 | 0.031 | 1 | | | |
| **Civil rights rate** | -0.181 | 0.285 | 0.214 | 0.236 | 0.048 | 0.020 | -0.066 | 1 | | |
| **Initial political rights** | 0.272 | -0.044 | 0.828 | -0.234 | 0.932 | 0.138 | 0.924 | -0.059 | 1 | |
| **Political rights rate** | -0.043 | 0.023 | -0.041 | 0.339 | -0.122 | 0.121 | -0.091 | 0.54 | -0.253 | 1 |

If there is any correlation between their wealth and a change in the quality of economic institutions, it is that improving economic institutions during an autocrat's rule is associated with higher autocrat wealth. These results are inconsistent with the hypothesis that autocrats maintain low-quality institutions because they benefit from them. The empirical results have implications with regard to both economic institutions and political institutions, which are discussed in the next two sections.

## 5. Autocrat Wealth and Economic Institutions

The paper's introduction notes the conventional wisdom that autocrats in countries with low-quality economic institutions do not improve those institutions because the political elite benefit from the low-quality institutions even though those institutions lower the economic well-being of the general population. Empirical evidence for that hypothesis is anecdotal, and an important question is: If the political elite benefit from maintaining low-quality economic institutions, what benefit does it receive? The only articles that look at a cross-section of countries to evaluate that hypothesis do not support it. Holcombe and Rodet (2012) find that the political elite can increase the resources under its control by improving the quality of economic institutions, and Holcombe and Boudreaux (2013) find that improving the quality of economic institutions lengthens the tenure of autocrats and stabilizes their political power. This paper finds that if anything, improving the quality of economic institutions is associated with higher autocrat wealth, and there is no evidence in any of these studies of any benefit to the political elite from maintaining low-quality economic institutions.

The evidence in this paper is based on a limited dataset, and certainly is not conclusive, but it does contribute some additional evidence to the literature that finds no benefit to autocrats from maintaining low-quality institutions, and finds that if anything, autocrats can improve their well-being in several dimensions by overseeing institutional improvements. Applied narrowly to the conventional wisdom, the question raised by this literature is, what is the benefit the political elite receive from maintaining low-quality institutions? Perhaps different data or empirical methods will reveal some benefit, but at this point, the literature has identified no benefit to the political elite from maintaining low-quality economic institutions. There is room for additional research on this question, either to identify the benefit the political elite gains from maintaining low-quality economic institutions, or to provide additional evidence supporting the studies that, at this point, all indicate that the political elite do not benefit from keeping their countries poor.

This empirical result raises the question of why, if autocrats would benefit personally from improving institutional quality, they do not make those improvements. One would think that autocrats would know their own interests at least as well as academics. One reason might be that institutional change is difficult to implement, so even if autocrats see the benefit from doing so, they may not be able to actually accomplish institutional reforms. Some autocrats—Augusto Pinochet and Park Chung-hee were mentioned earlier—have done so, while others have not. The idea that autocrats see advantages to institutional reform but are unable to accomplish it is a conjecture that goes beyond the analysis in this paper, so is a topic for future research. With regard to economic institutions, the major question this paper prompts is what benefit, if any, autocrats receive from maintaining low-quality institutions. Thus far, the literature has provided no systematic evidence that there are any benefits.

## 6. Autocrat Wealth and Political Institutions

The empirical results in this paper consistently find a strong negative correlation between the quality of political institutions at the time an autocrat assumes power and the wealth of autocrats. Wealthier autocrats tend to come to power in those countries in which the political institutions are poorest. A possible explanation for this finding is that in countries with low-quality political institutions, wealthy individuals seek political power to protect their wealth. Winters (2011: 7) notes, "In systems where property is reliably defended externally (especially by an armed state through institutions and strong property rights and norms), oligarchs have no compelling need to be armed or engaged directly in political roles." Bill Gates and Warren Buffett do not need to seek direct political power (although they do lobby to attempt to turn political outcomes their way) because political institutions in the United States protect their wealth, but in countries with lower-quality political institutions, Winters argues that the economic elite has more of an incentive to secure political power directly to prevent the political elite from appropriating its wealth.[6]

This suggests that the conventional wisdom may be based on a mistaken interpretation of the observation that political leaders in countries with low-quality institutions tend to be wealthy.

---

[6] Winters (2011: 6) defines oligarchs as "…actors who command and control massive concentrations of material resources that can be deployed to defend or enhance their personal wealth and exclusive social position." An example that illustrates the benefit of the wealthy assuming political power to protect their wealth is Mikhail Khodorkovsky of Russia, whose oil wealth at one time made him Russia's wealthiest individual, but was jailed for 10 years by Putin and had his assets appropriated by the state.

The observation of wealthy autocrats ruling over poorer countries suggests the conventional wisdom that those autocrats benefit from the low-quality institutions, but another possibility is that the wealthy seek political power to protect their wealth from predation by the political elite in countries that have a poor record of protecting property rights and maintaining rule of law. The elite does not benefit from maintaining low-quality institutions, but rather seeks political power to protect itself from being the victim of political predation under those institutions. Where political institutions are weak, wealth is more at risk to be confiscated by the political elite, so the wealthy have an incentive to convert some of its wealth into political power, to protect the rest.

We do not doubt that once it gains political power, the political elite will use that power to further its own interests, but examples of corrupt autocrats using political power to their advantage do not speak directly to the conventional wisdom. The question is whether the political elite would benefit more from maintaining low-quality institutions or from initiating institutional improvements, and all of the evidence so far—including the evidence in this paper—indicates that the political elite would be better off by overseeing improvements in institutional quality.

## 7. Conclusions

A substantial literature demonstrates that states with economic institutions that protect property rights and support market exchange prosper, while those with poor institutions that inhibit market activity remain poor. Mokyr (1990) and Landes (1998) give persuasive historical evidence, and Olson (1996) notes that profit opportunities do not remain unexploited for long. Poor countries remain that way because their economic institutions prevent innovative and entrepreneurial acts from being profitable. Entrepreneurial individuals are led to predatory rather than productive activity in countries with low-quality economic institutions, Baumol (1990) notes. Gwartney, Lawson, and Hall (2012) identify and quantify those institutions that lead to prosperity. If economists know what economic institutions lay a foundation for growth and prosperity, an important question for world economic development is why poor countries do not adopt those institutions. A commonly-given answer is that even though most of the population would benefit from institutional reform, the political elite benefits from maintaining those low-quality institutions, even as they impose costs on most people in their countries.

This paper has two main conclusions. First, it offers support for a small empirical literature which finds that autocrats do not benefit from maintaining low-quality institutions. Autocrats who maintain low-quality institutions do not increase their wealth as a result, and this finding is consistent with an existing literature finding that if anything, autocrats will be better off if they oversee improvements in institutional quality. The empirical evidence indicates that the hypothesis that autocrats in poor countries maintain low-quality political institutions because they benefit is incorrect.

Second, the paper finds that there is a negative correlation between the quality of political institutions and the wealth of autocrats at the time when they assume power. Rather than asking why autocrats maintain low-quality institutions, this result suggests a different question, which is why countries with lower-quality political institutions attract wealthier autocrats. A possible explanation for this is that high-quality political institutions protect individual wealth, so wealthy individuals in countries with high-quality institutions do not need to assume positions of political power to retain their economic power. Wealthy autocrats tend to assume power in countries with low-quality political institutions. They can use their wealth to try to buy political power, and then use their political power to try to protect and enhance their wealth. An important question is whether maintaining low-quality economic institutions furthers those goals, and this paper supports a recent literature that says it does not. The results in this paper go a good distance toward understanding why even though wealthy autocrats rule countries with poor institutions, they do not benefit from maintaining low-quality institutions.

When wealthy autocrats gain political power, they will use it to enhance and protect their wealth. We do not disagree with the conventional wisdom on this point. The question is whether it is to their advantage to maintain low-quality economic institutions, and this paper provides additional evidence to support the existing literature which says they do not. Political elites tend to be wealthier in countries with low-quality institutions not because they benefit from keeping institutional quality low, but because where institutional quality is low the wealthy have a greater incentive to seek political power.

Any empirical evidence on a hypothesis is always tentative. The questions addressed here are important for economic development and to date, the small amount of empirical research has not found evidence that low-quality institutions benefit the political elite. More research would be welcome, either to reinforce the tentative conclusion that the political elite does not benefit from maintaining low-quality economic institutions, or to provide evidence that when other data are examined or other methods are used, there is indeed evidence to support the hypothesis that autocrats receive personal benefits from maintaining these institutions. It appears that autocrats in countries with low-quality institutions tend to be wealthier because the wealthy have a greater incentive to seek political power in those countries, to protect their wealth, but this tentative conclusion would benefit from further research.

Important implications for development policy turn on the question. One would be hard-pressed to argue that autocrats do not understand what is in their own interest. If the conclusion that autocrats do not benefit from maintaining low-quality economic institutions holds up to further scrutiny, this would suggest that outside agencies and foreign governments can look for ways to help autocrats implement institutional reform. If the conventional wisdom is correct, then there is more of an argument for pushing for regime change and perhaps broader political reforms. Further research on this issue would be very worthwhile, regardless of whether it supports our finding or, after further study, finds contrary evidence.

**Appendix**

*1. Sources of leaders' wealth*

| Leader | Wealth Source |
| --- | --- |
| Jose Eduardo Dos Santos | http://docteurmilando.ivoire-blog.com/archive/2011/01/14/angola-jose-eduardo-dos-santos-possede-une-fortune-de-31-mil.html |
| Ilham Aliyev | http://www.celebritynetworth.com/richest-politicians/presidents/ilham-aliyev-net-worth/ |
| Hamad ibn isa al Khalifa | http://www.almanachdegotha.org/id229.html |
| Khaleda Zia | http://www.thedailystar.net/newDesign/news-details.php?nid=71084%3Cbr%20/%3E |
| Paul Biya | http://www.therichest.org/nation/richest-person-in-cameroon/ |
| wen jiabao | http://www.guardian.co.uk/news/blog/2012/oct/26/who-are-the-wealthiest-world-leaders |
| Juan Manuel Santos | http://en.mediamass.net/people/juan-manuel-santos/highest-paid.html |
| Mobutu Sese Seko | http://www.ghanafilla.net/archives/33753#.UbXY6fnqkfV |
| Hosni Mubarek | http://in.finance.yahoo.com/photos/8-of-the-richest-dictators-in-history-slideshow/hosni-mubarak-photo-1348740984.html |
| Girma Wolde-Giorgis | http://www.celebritynetworth.com/richest-politicians/presidents/girma-woldegiorgis-net-worth/ |
| Ali Bongo Ondimba | http://in.finance.yahoo.com/photos/8-of-the-richest-dictators-in-history-slideshow/ali-bongo-ondimba-photo-1348740985.html |
| mohamed suharto | http://www.cbv.ns.ca/dictator/Suharto.html |
| Ali Khamenei | http://money.ca.msn.com/savings-debt/gallery/dictators-and-their-wealth?page=5 |
| Uhuru Kenyatta | http://www.nairobiwire.com/2013/04/uhuru-kenyatta-is-africas-richest.html |
| H.H Sheikh Sabah IV Al-Ahmad Al-Jaber Al-Sabah | http://www.almanachdegotha.org/id229.html |
| Filip Vujanović | https://reportingproject.net/underground/index.php?option=com_content&view=article&id=6:ukanovis-montenegro-a-family-business&catid=3:stories&Itemid=19 |

| Leader | Wealth Source |
| --- | --- |
| Mohammad IV | http://www.therichest.com/celebnetworth/politician/royal/king-mohammed-iv-net-worth/ |
| Beatrix | http://www.forbes.com/sites/investopedia/2011/04/29/the-worlds-richest-royals/ |
| Goodluck Jonathan | http://nigeriavillagesquare.com/articles/ogaga-ifowodo/vice-president-jonathans-qmodestq-fortune.html |
| Harald V | http://e24.no/makro-og-politikk/kongens-formue-under-100-mill/3603222 |
| Qaboos Bin Said Al Said | http://www.almanachdegotha.org/id229.html |
| Asif Ali Zardari | http://www.daily.pk/president-asif-ali-zardari-2nd-most-richest-man-of-pakistan-6666/ |
| Benigno "Noynoy" Aquino III | http://www.affordablecebu.com/load/finance_wealth/president_benigno_quot_noynoy_quot_aquino_iii_net_worth/34-1-0-2692 |
| Hamad bin Khalifa | http://www.almanachdegotha.org/id229.html |
| Abdullah Bin Abdulaziz | http://www.forbes.com/2006/05/03/cz_forbes_0522_royals_slide.html?thisSpeed=35000 |
| Carl XVI Gustaf | http://exploredia.com/richest-politicians-in-the-world/ |
| Bashar Al-Assad | http://in.finance.yahoo.com/photos/8-of-the-richest-dictators-in-history-slideshow/bashar-al-assad-photo-1348740985.html |
| Bhumibol Adulyadej | http://www.almanachdegotha.org/id229.html |
| Zine El Abidine Ben Ali | http://money.ca.msn.com/savings-debt/gallery/dictators-and-their-wealth?page=3 |
| Sheikh Khalifa bin Zayed Al Nahyan | http://www.forbes.com/2006/05/03/cz_forbes_0522_royals_slide_3.html?thisSpeed=35000 |
| Yoweri Musevini | http://ddungu.wordpress.com/2012/04/13/president-yoweri-mueveni-s-accumulation-of-wealth-surprises-ugandans%E2%80%8F/ |
| Elizabeth II | http://www.forbes.com/sites/investopedia/2011/04/29/the-worlds-richest-royals/ |
| Robert Mugabe | http://in.finance.yahoo.com/photos/8-of-the-richest-dictators-in-history-slideshow/robert-mugabe-photo-1348740999.html |

*2. Definition of variables*

| Variable | Definition | Source |
|---|---|---|
| Log Wealth | Natural log of real wealth | See Appendix Table 1 |
| EFW rate | (EFW at end - EFW at beginning )/ tenure | Economic Freedom of the World 2012 |
| Initial EFW | EFW at beginning of tenure | Economic Freedom of the World 2012 |
| PCI rate | (PCI at end - PCI at beginning) / tenure | GDP per capita, World Bank, PPP |
| Initial PCI | PCI at beginning of tenure | GDP per capita, World Bank, PPP |
| Freedom Rate | (Freedom at end - Freedom at beginning)/ tenure | Freedom House, freedom rating |
| Initial Freedom | Freedom at beginning of tenure | Freedom House, freedom rating |
| Civil Rights Rate | (Civil Rights at end - Civil Rights at beginning)/ tenure | Freedom House, civil liberties |
| Initial Civil Rights | Civil Rights at beginning of tenure | Freedom House, civil liberties |
| Political Rights Rate | (Political Rights at end - Political Rights at beginning)/ tenure | Freedom House, political rights |
| Initial Political Rights | Political Rights at beginning of tenure | Freedom House, political rights |
| Polity2 Rate | (Polity2 at end - Polity2 at beginning)/ tenure | Polity IV |
| Initial Polity2 | Polity2 at beginning of tenure | Polity IV |
| Distance | Distance from the Equator | http://lab.lmnixon.org/4th/worldcapitals.html; http://www.eaae-astronomy.org/eratosthenes/ index.php?option=com_content&view=article &id=47&item&id-68 |
| Resource Rents | Natural resources (% of GDP) | Total natural resources, World Bank |
| ODA | Official Development Assistance (% of GDP) | IndexMundi.com, Net Official Development Assistance and Official Aid Received |
| Tenure | Length of Tenure (in years) | |